\documentstyle[aps,twocolumn]{revtex}
\begin{document}
\preprint{USACH-FM-00-06}
\title{Non-Commutative Quantum Mechanics}
\author{ J. Gamboa$^1$\thanks{E-mail: jgamboa@lauca.usach.cl}, M. Loewe
$^{2}$
\thanks{E-mail:
mloewe@fis.puc.cl,} and
J. C. Rojas$^3$\thanks{E-mail: rojas@sonia.ecm.ub.es}}
\address{$^1$Departamento de F\'{\i}sica, Universidad de Santiago de Chile,
Casilla 307, Santiago 2, Chile\\
$^2$ Facultad  de F\'{\i}sica, Pontificia Universidad  Cat\'olica de Chile,
Casilla 306, Santiago 22, Chile
\\
$^3$ Departament ECM, Facultat de Fisica, Universitat de Barcelona and Institut
D'Altes Energies,
\\
Diagonal 647, E-08028, Barcelona, Spain}
\maketitle
\begin{abstract}
A general non-commutative quantum mechanical system in a central potential
$V=V(r)$
in two dimensions  is considered. The spectrum is bounded from below
and for large values of the anticommutative parameter $\theta $,
we find an explicit expression for the eigenvalues.
In fact, any quantum mechanical system with these characteristics is equivalent
to a commutative one in such a way that the interaction $V(r)$ is replaced by
$V = V ({\hat H}_{HO}, {\hat L}_z)$, where
${\hat H}_{HO}$ is the hamiltonian of the two-dimensional harmonic oscillator
and ${\hat L}_z$ is z-
component of the angular momentum.
For other finite values of $\theta$ the model can be solved by using perturbation
theory.
\end{abstract}
\pacs{ PACS numbers:03.65.-w, 03.65.Db}

Recent results in string theory \cite{string} suggest that the spacetime could be
non-conmutative \cite{various}. This intriguing
possibility implies new and deep changes in our conception of spacetime that
could be visualized at the quantum mechanical level.
For example, unitarity in quantum mechanics is assured if time is commutative,
but  the spatial non-commutativity , although it is
completely consistent with the standard rules of quantum mechanics, imply the
new Heisenberg relation
\begin{equation}
\Delta x \,\Delta y \sim \theta, \label{1}
\end{equation}
where $\theta$  is the strength of the non-commutative effects and plays an
analogous role to
$\hbar$ in usual quantum mechanics.

In this letter we would like to discuss a general non-commutative quantum
mechanical
system  stressing the differences with the equivalent commutative case.
More precisely, we show that any two-dimensional non-commutative system in a
central potential
$V = V(r)$ where $r = \sqrt{\vert {\bf x}\vert^2}$
is equivalent to a commutative system decribed by the potential
\begin{equation}
V = V ({\hat H}_{HO}, {\hat L}_z), \label{2}
\end{equation}
where ${\hat H}_{HO}$ is the hamiltonian of the usual
(commutative) two-dimensional harmonic oscillator and
${\hat L}_z$ is the z-component of the angular momentum.

In the non-commutative space  one replace the ordinary product by the Moyal or
star product
\begin{equation}
{\bf A}\star {\bf B}({\bf x}) = e^{\frac{i}{2}\theta^{ij}\partial^{(1)}_i \partial^{(2)}_j}
{\bf A}({\bf x}_1) {\bf
B}({\bf x}_2) \vert_{{\bf x}_1={\bf x}_2={\bf x}}. \label{4}
\end{equation}

The only modification to the Schr\"{o}dinger equation
\begin{equation}
i \frac{\partial \Psi ({\bf x},t)}{\partial t} = \biggl[ \frac{{\bf p}^2}{2 m} + V ({\bf x})
\biggl] \Psi ({\bf x},t),
\label{5}
\end{equation}
is to replace
\begin{equation}
V({\bf x})  \star \Psi ({\bf x},t)\rightarrow V({\bf x} -  \frac{{\tilde {\bf p}}}{2})\Psi , \label
{b}
\end{equation}
where $\tilde p _{i_k}=\theta^{i_kj_k} p_{j_k}$, being
$\theta _{ij} = \theta \epsilon _{ij}$
with $\epsilon _{ij}$ the antisimetric tensor.
This formula that appeared lately in connection with string
theory was written in \cite{sussk} although there is an older version
also  known as Bopp's shift \cite{bopp}.

The next step is to consider a central potential in two dimensions. The right
hand side of (\ref{b})
becomes
\begin{eqnarray}
V(\vert {\bf x} -  \frac{{\tilde {\bf p}}}{2} \vert^2)\Psi&=& V ( \frac{\theta^2}{4} p_x^2 +
x^2 +
\frac{\theta^2}
{4} p_y^2 + y^2  - \theta L_z)\Psi \nonumber
\\
&=& V ({\hat {\aleph}})\Psi
, \label{6}
\end{eqnarray}
where the ${\hat \aleph}$ operator is defined as
\begin{equation}
{\hat \aleph} = {\hat H}_{HO} - \theta{\hat L}_z, \label{7}
\end{equation}
and corresponds to  a two-dimensional
harmonic oscillator with effective mass $m= 2/\theta^2$,  frequency
$\omega = \theta$ and angular
momentum $L_z = x p_y - y p_x$. The symmetry group for this system is $SU(2)$
and
the spectrum of ${\hat \aleph}$ can be computed noticing that
\begin{eqnarray}
L_x &=& \frac{1}{2} ( a^{\dagger}_x a_x - a^{\dagger}_y a_y), \nonumber
\\
L_y &=& \frac{1}{2} ( a^{\dagger}_x a_y + a^{\dagger}_y a_x), \label{8}
\\
L_z &=& \frac{1}{2i} ( a^{\dagger}_x a_y - a^{\dagger}_y a_x), \nonumber
\end{eqnarray}
are symmetry generators satisfying the Lie algebra $[L_i,L_j] = i\epsilon_{ijk}L_k$
and,
therefore, $\{{\hat
\aleph}, {\bf \hat L}^2, J_z = \frac{1}{2}{\hat L}_z\}$ is a complete set of conmuting
observables. If we
denote by  $\lambda_{jm}$
and $\vert j,m>$ the eigenvalues and eigenvectors, respectively, then we have
the
selection rules
\begin{eqnarray}
j&=& 0, \frac{1}{2}, 1, \frac{3}{2}, ...\nonumber
\\
m&=& j, j-1, j-2, ..., -j . \label{selec}
\end{eqnarray}
The eigenfunctions $\vert j,m>$ are well known\cite{Baym} and the eigenvalues of
${\hat \aleph}$
are given by
\begin{equation}
\lambda_{jm} = \theta \,[\,2j + 1- 2m].\label{eigen}
\end{equation}

Using these results,  the calculation of the eigenvalues of $V({\hat \aleph})$ is
straightforward. Indeed, if the
eigenvalues of the operator ${\hat A}$ are $ a_n $, then the function $f({\hat A})$,
after
expanding for small values
of $\epsilon$,  is
\begin{eqnarray}
f(\hat{A} + \epsilon) \psi_n &=&
\left( f(\hat{A}) + f^{\prime}(\hat{A}) \epsilon
+ \frac{1}{2!} f''(\hat{A}) \epsilon ^2 \cdots
\right) \psi_n
\nonumber \\
&=&
\left( f(a_{n}) + f^{\prime}(a_{n}) \epsilon
+ \frac{1}{2!} f''(a_{n}) \epsilon ^2 \cdots
\right) \psi_n
\nonumber \\
&=& f(a_n + \epsilon)\psi_n  \rightarrow
f(a_n )\psi_n,
\label{fatil}
\end{eqnarray}
and, as a consequence, the eigenvalue equation of $V({\hat \aleph})$ is
\begin{equation}
V(\hat{\aleph}) \mid j,m > = V\left[\,\theta \,(2j+1-2m) \right] \mid j,m>. \label{aleph}
\end{equation}
 Once  equation (\ref{aleph}) is found, one must compute the spectrum of the  full
 hamiltonian given by
 \begin{eqnarray}
 H &=& \frac{{\bf p}^2}{2M}
 + V(\hat{\aleph})
 \nonumber \\
 &=&
 \frac{2}{M \theta^2} \left(
 \frac{\theta^2 }{4} {\bf p}^2
 + {\bf r}^2 -\theta L_z
 \right) - \frac{2}{M \theta^2} {{\bf r}^2}
 + V(\hat{\aleph}) + \frac{2}{M\theta} L_z
 \nonumber \\
 &=& \frac{2}{M \theta^2} \hat{\aleph}+ V(\hat{\aleph})
 - \frac{2}{M \theta^2} {\bf r}^2 +\frac{2}{M\theta} L_z
 \nonumber \\
 &\equiv& H_0 - \frac{2}{M \theta^2} {\bf r}^2 + \frac{2}{M\theta} L_z. \label{ori}
 \end{eqnarray}

By using equations (\ref{eigen}) and (\ref{aleph}) one find that the eigenvalues  of
${\hat
H}_0$ are
\begin{equation}
\Lambda_{j,m} = \frac{2}{M \theta} \left[ 2j+1-2m \right]
+ V \left[ \,\theta \, (2j+1-2m)\right].
\end{equation}

The second term of the Hamiltonian can be treated as a perturbation for
large (but finite) values of $\theta $.  For regular, polynomial-like
potentials, the situation is similar to the solitonic case as in \cite{stromin}.

We will concentrate on the expectations values
of the full Hamiltonian, $E_{jm} = \langle j,m | \hat{H} | j,m \rangle$  \it i.e. \rm

\begin{eqnarray}
E_{jm}& = & <j,m\vert \hat{H}_0 \vert j,m> - \frac{2}{M \theta^2}<j,m\vert {\bf r}^2
\vert
j,m>  \nonumber
\\
&+&  \frac{4}{M \theta}<j,m\vert L_z\vert
j,m>, \nonumber
\\
&=& \frac{2}{M \theta} [ 2 j +1 ] + V[\, \theta \,(2j + 1- 2m)] \nonumber
\\
& -& \frac{2}{M \theta^2}  < j,m \mid  { {\bf r}^2}  \mid j,m> .
\label{formula}
\end{eqnarray}

The next to last term in the right hand side in (\ref{formula}) can
be calculated using perturbation theory for  large values
of  $\theta$. Indeed, in such case $\vert j,m>$
corresponds to the two-dimensional harmonic oscillator eigenvectors
\begin{equation}
\vert j,m> = \frac{{a_+^\dagger}^{j+m} {a_-^\dagger}^{j-m}}{\sqrt{(j+m)! (j-m)!}}
\vert 0,0>, \label{os}
\end{equation}
\noindent
where  in (\ref{os}) we have used the Schwinger representation  for the two-
dimensional
harmonic oscillator \cite{Baym}
and $\,\,\,\,\,$ $ <j,m\mid {\bf r}^2 \mid j,m>$ becomes

\begin{equation}
<j,m\mid {\bf r}^2 \mid j,m> = \frac{\theta}{2}[2j+1].
\end{equation}

\noindent
Let us consider now two kinds of  singular potentials.

a) If $V(r)= - \gamma/r^{\alpha}$, then
$ V(\theta)= - \gamma {[\theta (2j +1-2m)]}^{-\frac{\alpha}{2}}$.
Note that this term goes like $\theta ^{-\frac{\alpha }{2}}$, and therefore, the
relevant contribution to the spectrum,
for $\alpha > 2$ is given by the first term  in (\ref{formula}). In fact the difference
between the levels $2j$ and  $2j +
1$ is $2/M\theta$.

b) From the physical point of view, probably the most interesting case is the
Coulomb potential
 $V(r) = \gamma \ln(r)$ which corresponds to
 $V(\theta)=  \gamma /2 \ln[{\theta  (2j +1-2m)}]$.
 This could have a  relation to the Quantum Hall  Effect where electrons are
 confined in a plane. We would like to
 remark that spectroscopy  in two dimensional systems could be a sensible
 mechanism for detecting
 non-commutative corrections to quantum mechanics\cite{sheikh}.

 \noindent
Finally we would like to point out that our results seem to indicate that the
connection between  the
 commutative and non-commutative regimes is abrupt, {\it i.e.}  the limit
 $\theta \rightarrow 0$ cannot be taken
 directly\cite{min}.

\acknowledgements

We would like to thank V. O. Rivelles and F. M\'endez for several discussions on
non-commutative geometry. This work has
been partially supported by the grants Nr. 1980788 and Nr. 1980577 from
Fondecyt-Chile and Dicyt-USACH.


\begin{references}
\bibitem{string}  A. Connes, M. Douglas and A. S. Schwarz, {\it JHEP} 9802:003 (
1998); N. Seiberg and E. Witten, JHEP {\bf 09},
032 (1999); T. Yoneya, {\it  Prog.Theor.Phys.} {\bf 103}, 1081 (2000).
\bibitem{various} The literature is very extense, some references are; T. Filk, {\it
Phys.Lett.} {\bf B376}, 53 (1996); R. Gopakumar,
J. Maldacena, S. Minwalla and A. Strominger, {\it JHEP} {\bf 0006} 036 (2000); L.
Alvarez-Gaum\'e and S. Wadia, hep-
th/0006219; M. Hayakawa, {\it Phys. Lett.} {\bf 476}, 431 (2000); C. P. Martin and
F. Ruiz, hep-th/0007131; J.
Gomis and T. Mehen, hep-th/0005129; I. Mociou, M.
Popelov and R. Roibar, {\it Phys. Lett.} {\bf 489B}, 390 (2000);  J. Lukierski, P.
Stichel and W. Zakrzewski, {\it Ann
Phys.} {\bf 260}, 224 (1997); K. Valavano, hep-th/0006245; V.P. Nair and  A.P.
Polychronakos, hep-th/0011172.
\bibitem{sussk} D. Bigatti and L. Susskind, {\it Phys.Rev.} {\bf D62}, 066004
(2000); L. Mezincescu, {\it Star
Product in Quantum Mechanics}, hep-th/0007046; B. Morariu, A. P.
Polychronakos, hep-th/0102157.
\bibitem{bopp} see {\it e.g}, T. Curtright, D. Fairlie and C. Zachos  {\it Phys. Rev. }
{\it D58}, 025002 (1998); C.
Zachos,  {\it J. Math. Phys.}   {\bf 41}, (2000).
\bibitem{Baym} G. Baym, {\it Lectures on Quantum Mechanics},  pag. 380,
Benjamin (1974).
\bibitem{stromin} R. Gopakumar, S. Minwalla and A. Strominger, {\it JHEP} {\bf
05},020 (2000).
\bibitem{sheikh} In this context see; J. Gamboa, M. Loewe and J. C. Rojas, hep-th
/0101081; D. Bak , S. K. Kim, K. Su Soh, J. H.
Yee,  {\it Phys. Rev. Lett.}  {\bf 85}, 3087 (2000); {\it ibid}
hep-th/0006087
\bibitem{min} S. Minwalla, M. Van Raamsdonk and N. Seiberg, hep-th/9912072 .
\end{references}
\end{document}